\documentclass[12pt]{article}
\usepackage{epsfig,graphicx,scicite}
\usepackage{times}
\newcommand{\VLII}{{\it Via Lactea II}}
\newcommand{\msun}{\,\rm M_\odot}

\newcommand{\be}{\begin{equation}}
\newcommand{\ee}{\end{equation}}
\newcommand{\ba}{\begin{eqnarray}}
\newcommand{\ea}{\end{eqnarray}}
\newcommand{\f}{\frac}

\newcommand{\Vmax}{V_{\rm max}}
\newcommand{\rVmax}{r_{\rm Vmax}}

\newcommand{\kms}{\,{\rm km\,s^{-1}}}

\newcommand{\mcut}{m_0}
\newcommand{\sigv}{\langle \sigma v \rangle}
\def\spose#1{\hbox to 0pt{#1\hss}}
\newcommand{\lta}{\mathrel{\spose{\lower 3pt\hbox{$\mathchar"218$}}
      \raise 2.0pt\hbox{$\mathchar"13C$}}}
\newcommand{\gta}{\mathrel{\spose{\lower 3pt\hbox{$\mathchar"218$}}
      \raise 2.0pt\hbox{$\mathchar"13E$}}}

\newcommand{\Fermi}{\textit{Fermi}}

\newcommand{\captionfontsize}{\footnotesize}

\newenvironment{sciabstract}{%
\begin{quote} \bf}
{\end{quote}}

\newcounter{lastnote}
\newenvironment{scilastnote}{%
\setcounter{lastnote}{\value{enumiv}}%
\addtocounter{lastnote}{+1}%
\begin{list}%
{\arabic{lastnote}.}
{\setlength{\leftmargin}{.22in}}
{\setlength{\labelsep}{.5em}}}
{\end{list}}

\title{Exploring Dark Matter with Milky Way substructure} 

\author
{Michael Kuhlen$^{1}$\footnote{To whom correspondence should be addressed; E-mail: mqk@ias.edu.},
Piero Madau$^{2}$, Joseph Silk$^{3}$\\
\\
\normalsize{$^1$School of Natural Sciences, Institute for Advanced Study,
Princeton, NJ 08540}\\
\normalsize{$^2$Department of Astronomy \& Astrophysics, University of California,
Santa Cruz, CA 95064}\\
\normalsize{$^3$Department of Physics, University of Oxford, Oxford, OX1 3RH, UK}
}

\date{}

\begin{document} 

\maketitle 


\begin{sciabstract}
  The unambiguous detection of Galactic dark matter annihilation would
  unravel one of the most outstanding puzzles in particle physics and
  cosmology. Recent observations have motivated models in which the
  annihilation rate is boosted by the Sommerfeld effect, a
  non-perturbative enhancement arising from a long range attractive
  force. Here we apply the Sommerfeld correction to \textit{Via Lactea
    II}, a high resolution \textit{N}-body simulation of a
  Milky-Way-size galaxy, to investigate the phase-space structure of
  the Galactic halo. We show that the annihilation luminosity from
  kinematically cold substructure can be enhanced by orders of
  magnitude relative to previous calculations, leading to the
  prediction of $\gamma$-ray fluxes from up to hundreds of dark clumps
  that should be detectable by the \Fermi\ satellite.
\end{sciabstract}



In the standard cold dark matter (CDM) paradigm of structure
formation, a weakly interacting massive particle (WIMP) of mass
$m_\chi\sim 100$ GeV-10 TeV ceases to annihilate when the universe
cools to a temperature of $T_f \sim m_\chi/20$, about one nano-second
after the Big Bang. A thermally-averaged cross-section at freeze-out
of $\langle \sigma v\rangle_0\approx 3\times 10^{-26}$ cm$^{3}$
s$^{-1}$ results in a relic abundance consistent with observations
\cite{Jungman1996}.  Perturbations in the dark matter density are
amplified by gravity after the universe becomes matter dominated,
around ten thousand years after the Big Bang: the smallest structures
(``halos'') collapse earlier when the universe is very dense and merge
to form larger and larger systems over time.  Today, galaxies like our
own Milky Way are embedded in massive, extended halos of dark matter
that are very lumpy, teeming with self-bound substructure
(``subhalos'') that survived this hierarchically assembly process
\cite{Klypin1999,Moore1999,Diemand2007}.  Indirect detection of high
energy antiparticles and $\gamma$-rays from dark matter halos provides
a potential ``smoking gun'' signature of WIMP annihilation
\cite{Silk1984}. The usual assumption that WIMP annihilation proceeds
at a rate that does not depend, in the non-relativistic $v/c\ll 1$
limit, on the particle relative velocities implies that the primary
astrophysical quantity determining the annihilation luminosity today
is the local density squared. WIMP annihilations still occur in the
cores of individual substructures, but with fluxes that are expected
to be dauntingly small. The latest calculations show that only a
handful of the most massive Galactic subhalos may, in the best case,
be detectable in $\gamma$-rays by the {\it Fermi} satellite
\cite{Kuhlen2008,Pieri2008}.

The Sommerfeld enhancement, a velocity-dependent mechanism that boosts
the dark matter annihilation cross-section over the standard $\langle
\sigma v\rangle_0$ value
\cite{Hisano2004,March-Russell2009,Lattanzi2009,Arkani-Hamed2009}, may
provide an explanation for the experimental results of the PAMELA
satellite reporting an increasing positron fraction in the local
cosmic ray flux at energies between 10 and 100 GeV \cite{Adriani2008},
as well as for the surprisingly large total electron and positron flux
measured by the ATIC and PPB-BETS balloon-borne experiments
\cite{Chang2008,Torii2008}.  Very recent Fermi \cite{Abdo2009} and
H.E.S.S. \cite{Aharonian2009} data appear to be inconsistent with the
ATIC and PPB-BETS measurements, but still exhibit departures with
respect to standard expectations from cosmic ray propagation models.
Although conventional astrophysical sources of high energy cosmic
rays, such as nearby pulsars or supernova remnants, may provide a
viable explanation \cite{Profumo2009,Shaviv2009,Malyshev2009}, the
possibility of Galactic DM annihilation as a source remains intriguing
\cite{Bergstrom2009,Meade2009b,Kuhlen2009}. In this case,
cross-sections a few orders of magnitude above what is expected for a
thermal WIMP are required \cite{Cirelli2009}.

\begin{figure}[htp]
\includegraphics[width=\textwidth]{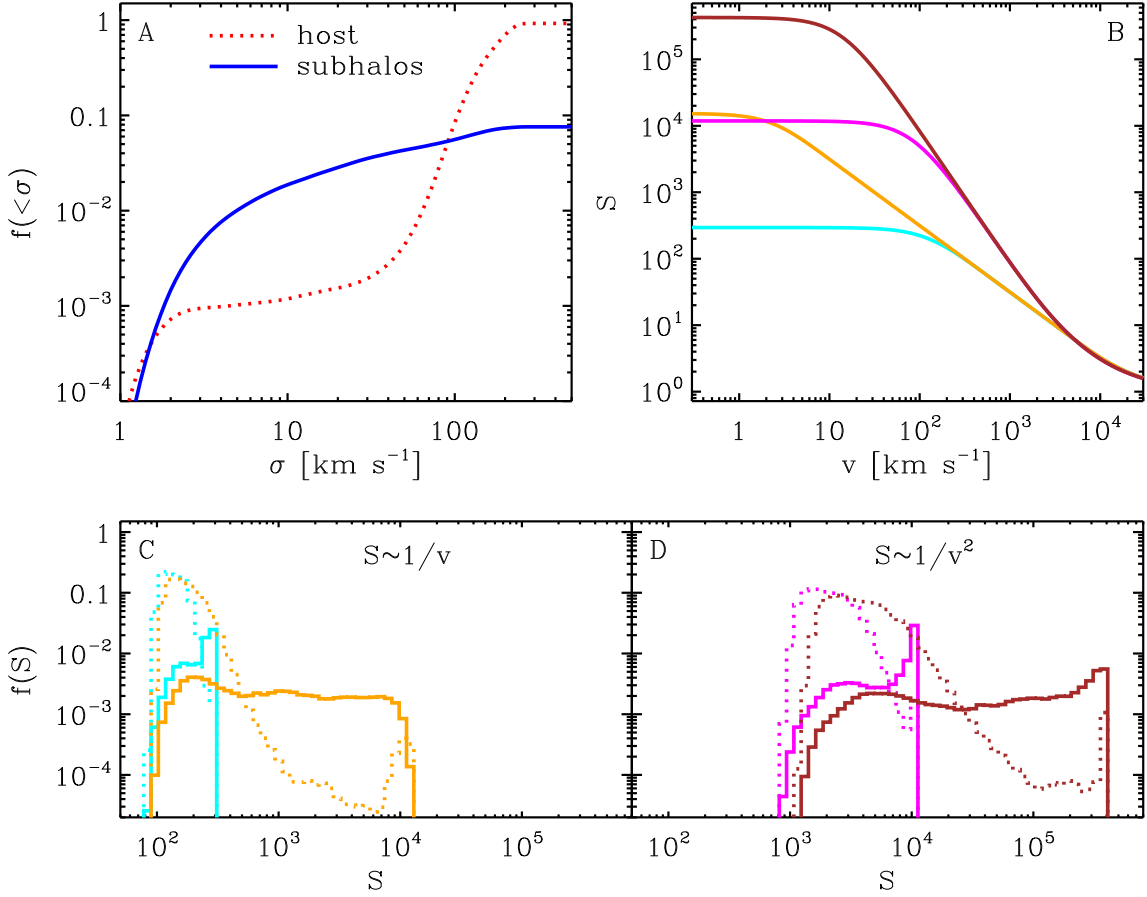}
\caption{\captionfontsize A: The distribution of velocity dispersion for
\VLII\ particles within 400 kpc. The dispersions are calculated from
the nearest 32 neighbors of each particle.  B: Sommerfeld enhancement
factor as a function of velocity for four representative models,
exhibiting $S \sim 1/v$ (cyan and orange curves) and $\sim 1/v^2$
behavior (magenta and brown), and high (cyan and magenta) versus low
(orange and brown) saturation velocities.  C and D: The corresponding
distributions of S-factors for \VLII\ particles.}
\end{figure}

The Sommerfeld non-perturbative increase in the annihilation
cross-section at low velocities is the result of a generic attractive
force between the incident dark matter particles that effectively
focuses incident plane-wave wavefunctions.  The force carrier may be
the $W$ or $Z$ boson of the weak interaction \cite{Lattanzi2009},
$m_\phi=80-90$ GeV/$c^2$, or a lighter boson, $m_\phi\sim$ GeV/$c^2$,
mediating a new interaction in the dark sector
\cite{note1,Arkani-Hamed2009}.  Upon introduction of a force with
coupling strength $\alpha$, the annihilation cross-section is shifted
to $\langle \sigma v\rangle=S\langle \sigma v\rangle_0$, where the
Sommerfeld correction $S$ disappears ($S=1$) in the limit
$v/c\rightarrow 1$ (thus leaving unchanged the weak scale annihilation
cross-section during WIMP freeze-out in the early universe). When
$v/c\ll \alpha$, $S \approx \pi\alpha c/v$ (``$1/v$'' enhancement),
but it levels off to $S_{\rm max}\approx 6\alpha m_\chi/m_\phi$ at
$v/c\approx 0.5 m_\phi/m_\chi$ because of the finite range of the
interaction. For specific parameter combinations, i.e. when
$m_\chi/m_\phi\approx n^2/\alpha$ where $n$ is an integer, the
(Yukawa) potential develops bound states, and these give rise to
large, resonant cross-section enhancements where $S$ grows
approximately as $1/v^2$ before saturating (see Supporting Online
Material).

\begin{figure}[htp]
\centering
\includegraphics[height=0.7\textheight]{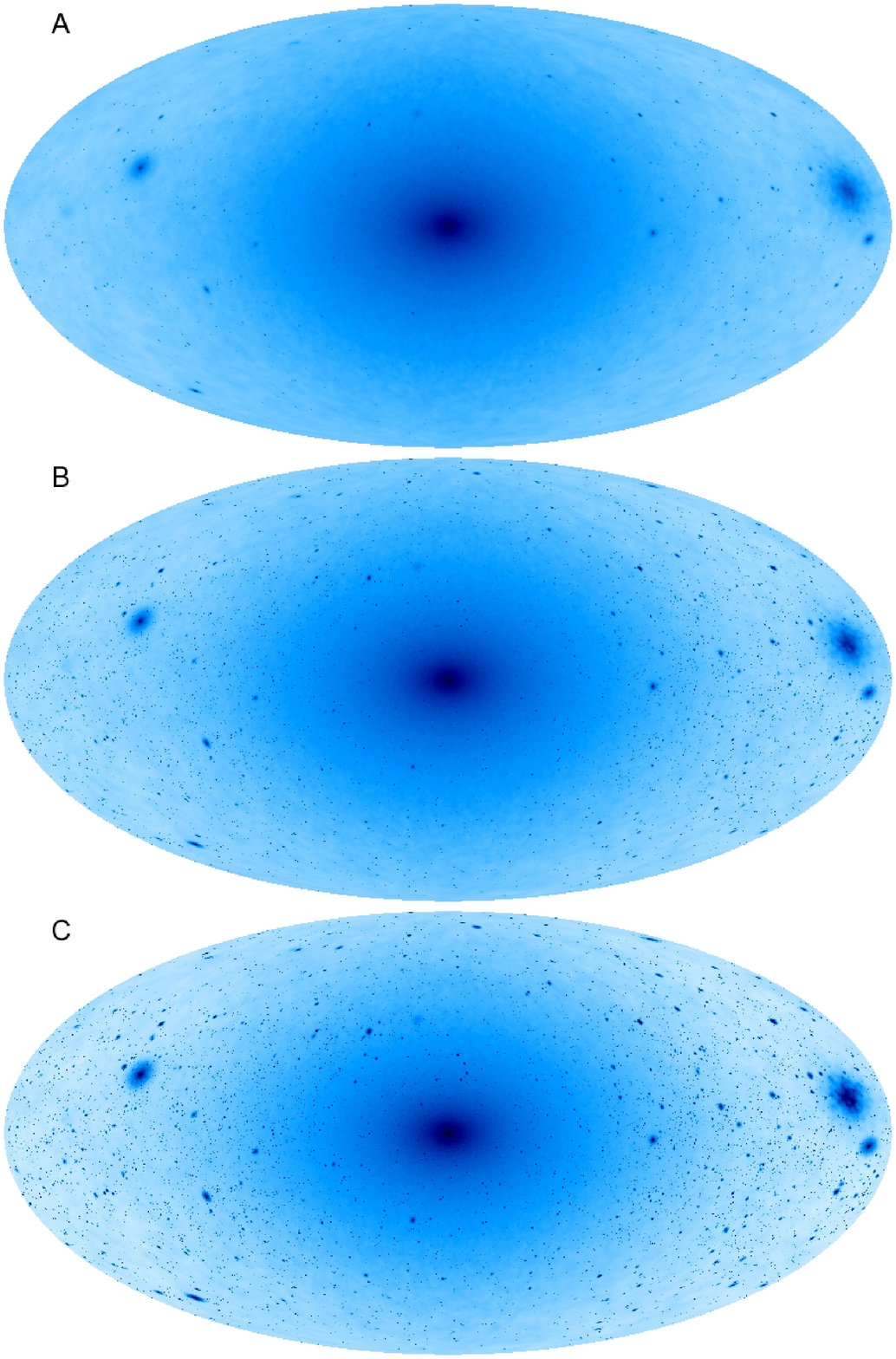}
\caption{\captionfontsize All-sky maps (in a Mollweide projection) of the
Sommerfeld-enhanced annihilation surface brightness ($\int_{\rm los}
\rho^2 S \; d\ell$) from all \VLII\ dark matter particles within 400
kpc. The observer is located at 8 kpc from the halo center along the
host halo's intermediate principal axis.  A: No Sommerfeld
enhancement.  B: $S \sim 1/v$, saturated at $\sim 1$ km s$^{-1}$. C:
$S \sim 1/v^2$ saturated at $\sim 5$ km s$^{-1}$. The maps have been
normalized to give the same total smooth host halo flux.}
\end{figure}

The Sommerfeld effect connects dynamically the dark and the
astrophysics sectors. Because the typical velocities of dark matter
particles in the Milky Way today are of the order of $v/c\sim
10^{-3}$, the resulting boost in the annihilation rate may provide an
explanation to the puzzling Galactic signals.  Compared to particles
in the smooth halo component, the Sommerfeld correction preferentially
enhances the annihilation luminosity of cold, lower velocity
dispersion substructure, as emphasized previously by
\cite{Lattanzi2009,Robertson2009,Bovy2009}. Detailed knowledge of the
full phase-space density of dark matter particles in the Milky Way is
thus necessary to reliably compute the expected signals.  Here we use
the \VLII\ cosmological simulation, a high precision calculation of
the assembly of the Galactic CDM halo, for a systematic investigation
of the impact of Sommerfeld-corrected models on present and future
indirect dark matter detection efforts.  \VLII\ employs just over one
billion $4,100\,\msun$ particles to follow, with a force resolution of
40 pc, the formation of a $1.9\times 10^{12}\,\msun$ Milky-Way size
halo and its substructure from redshift $z=104$ to the present
\cite{Diemand2008,Madau2008,Zemp2009}. (Fig.~1 A) The smooth halo
particles, whose velocity dispersions are set by the global potential,
typically have three-dimensional velocity dispersion $\sigma >
100\,\kms$.  Particles in self-bound subhalos dominate at lower
velocity dispersions. The total mass fraction of particles with
$\sigma<5\,\kms$ is 1\%. We calculated the Sommerfeld enhancement
factors $S$ on a particle-by-particle basis by averaging $S(v)$ over a
Maxwell-Boltzmann distribution of relative velocities with
one-dimensional velocity dispersion given by $\sqrt{2/3} \; \sigma$
(see the Supporting Online Materials for details) (Fig.~1 C, D).

The large Sommerfeld boost expected for $v/c\sim 10^{-4}-10^{-5}$ make
cold subhalos more promising sources of annihilation $\gamma$-rays
than the higher density but much hotter region around the Galactic
Center (Fig.~2). In Sommerfeld-enhanced models, substructures are much
more clearly visible, and can even outshine the Galactic Center when
the cross-section is close to resonance and saturates at low
velocities. Furthermore, baryonic processes will tend to heat up the
Galactic Center and dim its Sommerfeld boost, and thereby increase the
relative detectability of subhalos. Dark matter halos are not
isothermal and have smaller velocity dispersions in the center (see
Supporting Online Material).  In addition to an overall increase in
the annihilation rate, this ``temperature inversion'' leads to a
relative brightening of the center at the expense of the diffuse flux
from the surrounding region (Fig.~3). The subhalo exhibits its own
population of subclumps, also Sommerfeld-enhanced.

\begin{figure}[t]
\includegraphics[width=\textwidth]{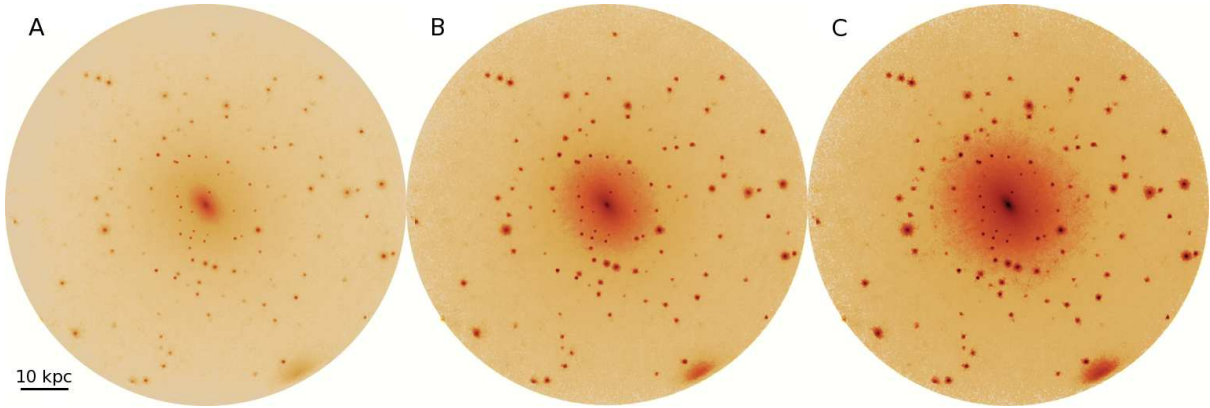}
\caption{\captionfontsize Annihilation rate maps (projections of $\rho^2S$
out to the tidal radius) of one of the most massive ($M \sim 2 \times
10^9\,\msun$) subhalos in \VLII, for the same models as in Fig.~2.
The images have not been normalized.  Compared to the $S=1$ case (A),
the total luminosity is 2,200 and 160,000 higher for $S\sim1/v$ (B)
and $S\sim1/v^2$ (C), respectively.}
\end{figure}
To address quantitatively the detectability of Sommerfeld-enhanced
subhalos by the \Fermi\ Space Telescope, we have converted the
annihilation flux calculated from our simulation \cite{note2} into a
predicted $\gamma$-ray flux and compared it to the expected
backgrounds. We investigated two different classes of particle physics
models (Table~1):

\begin{enumerate}
\item[i)] those motivated by \cite{Lattanzi2009}, in which the force
  carrier is the conventional weak force gauge boson, the W or Z
  particle, and the mass of the dark matter particle is $\gta 4$ TeV.
  We have chosen four representative values of $m_\chi$ and $\alpha$,
  which lie increasingly close to a $S \sim 1/v^2$ resonance. In these
  models the main source of $\gamma$-rays is the decay of neutral
  pions that are produced in the hadronization of the annihilation
  products;

\item[ii)] and those in which the annihilation is mediated by a new
  dark sector force carrier $\phi$ \cite{Arkani-Hamed2009}. The choice
  of parameters ($m_\chi, m_\phi, \alpha$) follows Meade, Papucci, \&
  Volansky (MPV) \cite{Meade2009} and satisfies constraints from
  recent H.E.S.S. measurements of the Galactic Center
  \cite{Aharonian2006a} and the Galactic Ridge \cite{Aharonian2006b},
  as well as the PAMELA measurement of the local positron fraction
  above 10 GeV \cite{Adriani2008} and the ATIC \cite{Chang2008} and
  PPB-BETS \cite{Torii2008} measurements of the total $(e^+ + e^-)$
  flux above 100 GeV. Models MPV-1 incorporate all three constraints
  and models MPV-2 only the H.E.S.S. and PAMELA data.  We considered
  models away from (a) and close to (b) resonance, thereby covering
  both $S \sim 1/v$ and $\sim 1/v^2$ behavior. The data favor a light
  force carrier, $m_\phi \approx 200$ MeV, and the $\gamma$-rays
  originate then as final state radiation (internal bremsstrahlung)
  accompanying the decay of the $\phi$'s into $e^+e^-$ pairs.
\end{enumerate}
The magnitude of the relativistic cross section was fixed to the
standard value of $\sigv_0 = 3 \times 10^{-26}$ cm$^{3}$ s$^{-1}$.
$\gamma$-ray spectra are shown in Figure~S2 in the Supporting Online
Material.

\begin{table}
\centering
\caption{\captionfontsize Summary of the models used to assess subhalo detectability with \Fermi. Particle 
  physics parameters are: $m_\chi$, the mass of the dark matter particle, $m_\phi$, the mass of the force carrier,
  and $\alpha$, the coupling constant. In the two right-most columns we give $S_{\rm max}$, the maximum Sommerfeld 
  enhancement obtained, and the saturation velocity $v_{\rm sat}$, defined as the velocity at which  
  $S$ reaches 90\% of $S_{\rm max}$. \medskip}
\begin{tabular}{c|ccc|rc}
 Model & $m_\chi$ & $m_\phi$ & $\alpha \times 100$ & $S_{\rm max}$ & $v_{\rm sat}$ \\
       & (TeV)   & (GeV)    &                     &              & (km s$^{-1}$) \\
\hline                                                                        
\hline                                                                        
  LS-1 & 4.30 &  90 & $3.307$ & 1,500 & 80 \\
  LS-2 & 4.45 &  90 & $3.297$ & 12,000 & 28 \\
  LS-3 & 4.50 &  90 & $3.288$ & 70,000 & 12 \\
  LS-4 & 4.55 &  90 & $3.281$ & 430,000 & 4.7 \\
\hline                                                                        
MPV-1a &  1.0 & 0.2 & $4.000$ & 3,000 & 7.4 \\
MPV-1b &  1.0 & 0.2 & $3.739$ & 16,000 & 2.4 \\
\hline                                                                       
MPV-2a & 0.25 & 0.2 & $4.000$ & 480 & 40 \\
MPV-2b & 0.25 & 0.2 & $4.500$ & 40,000 & 3.3 \\
\end{tabular}
\label{tab:models}
\end{table}

We determined the \Fermi\ detection significance by summing the
annihilation photons from all the pixels in our all-sky maps covering
a given subhalo, and compared this to the square root of the number of
background photons from the same area. We counted a subhalo as
``detectable'' if it had a total signal-to-noise greater than 5
(Table~\ref{tab:N5}). The numbers are quite large, implying that
individual subhalos should easily be detected by \Fermi\ if Sommerfeld
enhancements are important. Even in the most conservative cases
(MPV-1a and MPV-2a) around ten or more subhalos should be discovered
after 5 years of observation. In fact, on the basis of all models
considered here it is predicted that \Fermi\ should be able to
accumulate enough flux in its first year of observations to detect
several dark matter subhalos at more than $5 \sigma$ significance, a
prediction that will soon be tested, and that may open up the door to
studies of non-gravitational dark matter interactions and new particle
physics.  The central brightening discussed above results in a smaller
angular extent of a given subhalo's detectable region: the stronger
the Sommerfeld enhancement the fewer pixels exceed the detection
threshold.  Nevertheless, for all models considered here the majority
of detectable subhalos would be resolved sources for \Fermi.

\begin{table}
\centering
\caption{\captionfontsize Detectable Subhalos. The number of subhalos that would be detected with $> 5 \sigma$ 
significance by \Fermi\ after  1, 2, 5, and 10 years in orbit, for different Sommerfeld-enhanced dark 
matter particle models. In the two right-most columns we give the median distance and mass of the detectable 
clumps after 5 years in orbit. \medskip}
\begin{tabular}{c|cccc|cc}
 Model & 1 yr & 2 yr & 5 yr & 10 yr & $\tilde{D}$ & $\tilde{M}_{\rm sub}$\\
       &      &      &      &       & (kpc)       & ($\msun$) \\
\hline                                                                        
\hline                                                                        
  LS-1 &   12 &   19 &   29 &   38 & 24 & $1.4 \times 10^7$ \\
  LS-2 &   72 &   99 &  167 &  244 & 42 & $9.5 \times 10^6$ \\
  LS-3 &  225 &  311 &  457 &  583 & 56 & $6.2 \times 10^6$ \\
  LS-4 &  410 &  528 &  730 &  919 & 66 & $4.9 \times 10^6$ \\
MPV-1a &    5 &    7 &   12 &   15 & 16 & $9.8 \times 10^7$ \\
MPV-1b &    9 &   14 &   25 &   36 & 25 & $4.4 \times 10^6$ \\
MPV-2a &   12 &   18 &   29 &   38 & 24 & $1.4 \times 10^7$ \\
MPV-2b &  187 &  254 &  397 &  518 & 55 & $4.5 \times 10^6$ \\
\end{tabular}
\label{tab:N5}
\end{table}

Another question of interest is whether Sommerfeld-corrected
substructure would lead to a significant boost in the local production
of high energy positrons, arising from dark matter annihilation in
subhalos within a diffusion region of a few thousand parsecs from
Earth, as well as of antiprotons within a correspondingly larger
diffusion region.  The local dark matter distribution at the Sun's
location appears quite smooth in the highest-resolution numerical
simulations to date \cite{Diemand2008,Zemp2009,Vogelsberger2009}.
Tidal forces efficiently strip matter from subhalos passing close to
the Galactic Center and often completely destroy them. Further
substructure depletion may be expected from interactions with the
stellar disk and bulge.  In the \VLII\ simulation, the mean number of
$>10^5\,\msun$ subhalos within 1 kpc of the Sun is only 0.04, and one
must reach three times farther to find one clump on average. Without
the Sommerfeld effect, this dearth of nearby substructure leads to a
local annihilation boost of less than 1\%, and at most 20\% in the
rare case of a nearby clump, as found in a statistical approach
\cite{Lavalle2008}.  The picture changes with Sommerfeld enhancement.
The low velocity dispersion of cold substructure leads to a greatly
increased luminosity compared to the hotter smooth component. For
typical $1/v$ models, subhalos resolved in our simulation within 2 kpc
contribute on average about half as much luminosity as the smooth
component, and up to 5 times as much in rare cases. If the Sommerfeld
enhancement is resonant ($S \sim 1/v^2$), then these subhalos dominate
by a factor of 20 on average and by as much as 200 in rare cases.

\bibliography{scibib}

\begin{thebibliography}{}
\bibitem{note1} The ``dark sector'' of our universe consists of dark
  matter, dark energy, and possibly additional dark forces.
\bibitem{note2} Even at the extremely high resolution of \VLII\ the
  cuspy subhalo centers are often not sufficiently well resolved to
  give accurate central surface brightnesses. We corrected the centers
  according to an analytical model that has been calibrated to our
  numerical simulations (see Supporting Online Material).
\end{thebibliography}

\bibliographystyle{Science}

{}

\begin{scilastnote}
\item{Support for this work was provided by NASA through grant
    NNX08AV68G (P.M.) and by the William L. Loughlin Fellowship at the
    Institute for Advanced Study (M.K.). This work would not have been
    possible without the expertise and invaluable contributions of
    all the members of the Via Lactea Project team.  We thank M.
    Lattanzi, S.  Profumo, N. Arkani-Hamed, N. Weiner, P. Meade and T.
    Volansky for enlightening discussions, D. Shih for help with the
    Sommerfeld enhancement calculations, and M.  Papucci for providing
    $\gamma$-ray spectra.}
\end{scilastnote}

\noindent \textbf{Supporting Online Material} \\
www.sciencemag.org \\
Supporting text \\
Figs. S1, S2, S3, S4, S5, S6

\newpage

\renewcommand{\thefigure}{S\arabic{figure}}
\setcounter{figure}{0}

\renewcommand{\thesection}{S\arabic{section}}
\setcounter{section}{0}

\begin{center}
{\LARGE Supporting Online Material}

\medskip

{\large ``Exploring Dark Matter with Milky Way substructure''\\
Kuhlen, Madau, \& Silk}
\end{center}

\section{Sommerfeld enhancement}

\begin{figure}[htp]
\centering
\includegraphics[width=0.45\columnwidth]{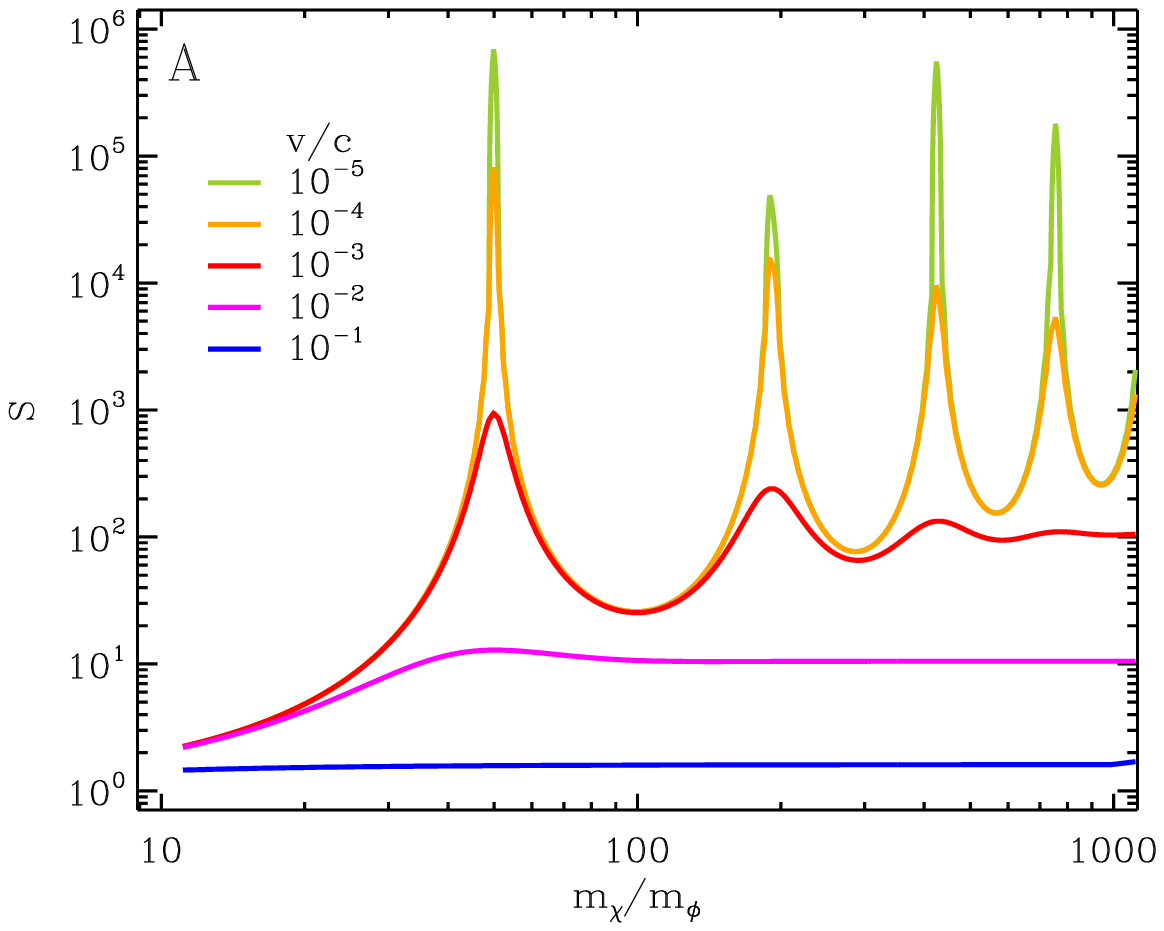}
\includegraphics[width=0.45\columnwidth]{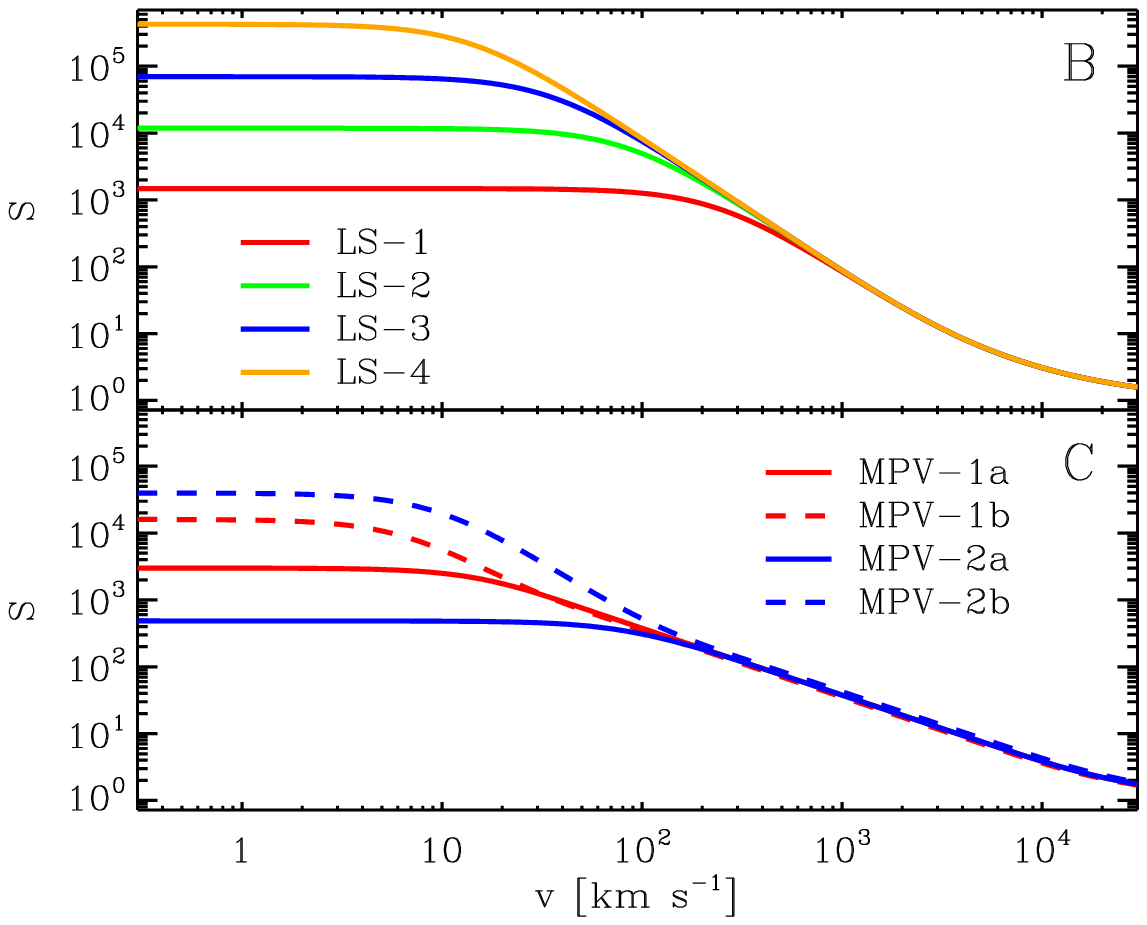}
\caption{\captionfontsize A: The Sommerfeld enhancement factor $S$ as
  a function of $m_\chi/m_\phi$ (the mass ratio of the dark matter
  particle to the force carrier particle) at a fixed coupling ($\alpha
  = 0.30$) for different velocities.  B and C: $S$ as a function of
  velocity for the models for which we calculate subhalo detectability
  with \Fermi. The parameters of the models are given in Table~1 in
  the main text.}
\label{fig:sommerfeld}
\end{figure}

A Sommerfeld enhancement to the annihilation cross-section arises when
the dark matter (DM) particle is heavy compared to the gauge boson
mediating the interaction: $m_\chi \gta m_\phi/\alpha$, where $\alpha
= \lambda^2/4\pi$, and $\lambda$ is the coupling between the dark
matter particle $\chi$ and the force carrier $\phi$. The magnitude of
the enhancement can be determined by solving the two-body radial
Schr\"odinger equation, as shown in
\cite{S_Arkani-Hamed2009,S_Lattanzi2009}. Defining the additional
dimensionless parameter
\begin{equation}
\beta^* = \sqrt{\f{\alpha m_\phi}{m_\chi}},
\end{equation}
one can distinguish three regimes for the dependence of the
enhancement $S$ as a function of the velocity \cite{S_Lattanzi2009}:
\begin{itemize}
\item[i)] for large velocities, $\beta \equiv v/c \gg \alpha$, there
  is no Sommerfeld enhancement, $S \sim 1$. This ensures that the
  relic abundance of dark matter is not affected, since the velocities
  were close to relativistic at freeze-out;
\item[ii)] at intermediate velocities, $\beta^* \ll \beta \ll \alpha$,
  the Sommerfeld enhancement follows a $1/v$ behavior, $S \approx \pi
  \alpha/\beta$;
\item[iii)] at low velocities, $\beta \ll \beta^*$, resonances appear
  for certain values of $m_\chi$, due to the presence of bound states.
  In this case $S$ grows as $1/v^2$ before saturating at a velocity
  that depends on how close to resonance $m_\chi$ lies. In the
  non-resonant case the Sommerfeld enhancement saturates when the
  deBroglie wavelength of the particle $\sim (m_\chi v)^{-1}$ becomes
  comparable to the range of the interaction $\sim m_\phi^{-1}$), i.e.
  when $v \sim m_\phi/m_\chi$.
\end{itemize}
Figure~\ref{fig:sommerfeld} shows the rich behavior of the Sommerfeld
enhancement, obtained by numerical integration of the Schr\"{o}dinger
equation with a Yukawa potential.

The magnitude of the Sommerfeld enhancement at a given location
depends on the details of the distribution of relative velocities of
the DM particles. Accurately determining the phase-space structure as
a function of position is a notoriously difficult task even with the
highest resolution N-body simulations. Some recent investigations
based on cosmological N-body simulations have found significant
structure in the coarse-grained velocity distribution of the host halo
and departures from a simple, single-''temperature'' Maxwell-Boltzmann
distribution \cite{S_Vogelsberger2009,S_Zemp2009}.  These features are
remnants of the accretion history of the host halo due to incomplete
phase mixing.  The centers of subhalos, however, are likely more
well-mixed and may show less significant departures from a Maxwellian.
At any rate, a full characterization of the phase-space distribution
at all locations in our simulation is beyond the scope of this paper,
and hence we make the simplifying assumption that the distribution of
relative velocities is of the Maxwell-Boltzmann form
\begin{equation}
  f(v_{\rm rel};\sigma_{\mu, \rm 1D}) = 4\pi \, \f{1}{(2\pi \sigma_{\mu, \rm 1D})^{3/2}} \, v_{\rm rel}^2 \, \exp\left[ - \f{1}{2} \left( \f{v_{\rm rel}}{\sigma_{\mu, \rm 1D}} \right)^2 \right],
\end{equation}
with a one-dimensional velocity dispersion
\begin{equation}
\sigma_{\mu, \rm 1D}^2 \equiv \f{k \, T}{\mu} = 2 \f{k \, T}{m} = 2 \,\sigma_{m, \rm 1D}^2,
\end{equation}
where $\mu = m_1 m_2/(m_1 + m_2) = m/2$ is the reduced mass of a two
particle system. From the simulated particles we determine a
three-dimensional velocity dispersion, $\sigma^2 \equiv \sigma_{m,\rm
  3D}^2 = 3 \, \sigma_{m,\rm 1D}^2 = 3/2 \, \sigma_{\mu, \rm 1D}^2$ (for
systems with zero velocity anisotropy). The Maxwell-Boltzmann-weighted
Sommerfeld enhancement is then given by
\begin{equation}
S(\sigma) = \int_0^\infty f(v;\sqrt{2/3} \,\sigma) \; S(v) \; dv.
\end{equation}

\section{Gamma-rays from dark matter annihilation}

\begin{figure}[htp]
\centering
\includegraphics[width=0.55\columnwidth]{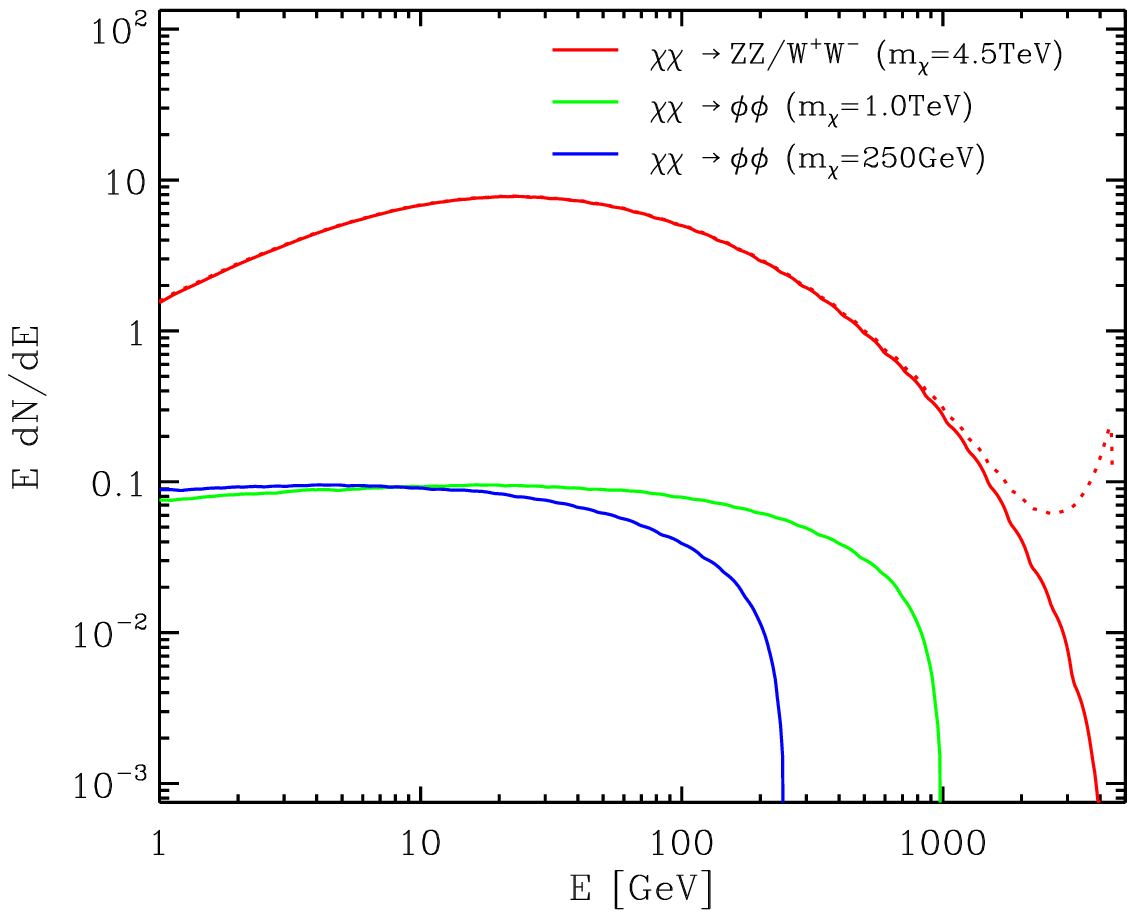}
\caption{\captionfontsize The $\gamma$-ray spectrum per annihilation
  for the models under consideration here. In the LS models ($\chi
  \chi \rightarrow Z Z \; {\rm or}\; W^+W^-$) the $\gamma$-rays come
  from the decay of pions produced in the decay of the bosons. The
  dotted line indicates the internal bremsstrahlung contribution in
  the case of $W^+W^-$ {\protect \cite{S_Bringmann2008}}. In the MPV
  models ($\chi \chi \rightarrow \phi \phi$) the $\gamma$-rays
  originate as final state radiation associated with the decay of the
  $\phi$ carriers into $e^+ e^-$ pairs.}
\label{fig:spectra}
\end{figure}

Since the dark matter particle is neutral it does not couple directly
to the electromagnetic field, and hence annihilations straight into
two monochromatic photons (or a photon and a Z boson) are typically
strongly suppressed.  Nevertheless $\gamma$-rays can be a significant
by-product of dark matter annihilations, since they can arise either
from the decay of neutral pions produced in the hadronization of the
annihilation products, or through internal bremsstrahlung associated
with annihilations into charged particles, or from interactions of
energetic leptons with the surrounding interstellar photons (inverse
Compton scattering). We do not consider the latter process here, since
we are focusing on the annihilation signal from dark matter subhalos
which are unlikely to harbor a sufficiently high stellar radiation
field.

The $\gamma$-ray spectra per annihilation that we use in our
detectability calculation are shown in Figure~\ref{fig:spectra}. In
the Lattanzi \& Silk models the annihilation results in two neutral
$Z$ bosons or a pair of $W^+$ and $W^-$ bosons, and the dominant
source of $\gamma$-rays is neutral pion decay. For $m_\chi=4.5$ TeV,
every annihilation results in $\sim 26$ photons with energies between
3 and 300 GeV. In the MPV models, the mass of the $\phi$ particle is
so low (by design), that only decays into $e^+ e^-$ pairs are
kinematically allowed, and we must rely on final state radiation
(internal bremsstrahlung) for the $\gamma$-ray signal. This results in
fewer 3-300 GeV $\gamma$-rays per annihilation (0.39 for $m_\chi=1$
TeV, 0.30 for $m_\chi=250$ GeV), but this is partially compensated by
the smaller dark matter particle mass and hence higher number density
at fixed mass density.

\section{The velocity structure of the Via Lactea II host and its
  subhalos}

\begin{figure}
\vspace*{-0.5in}
\centering
\includegraphics[width=0.49\textwidth]{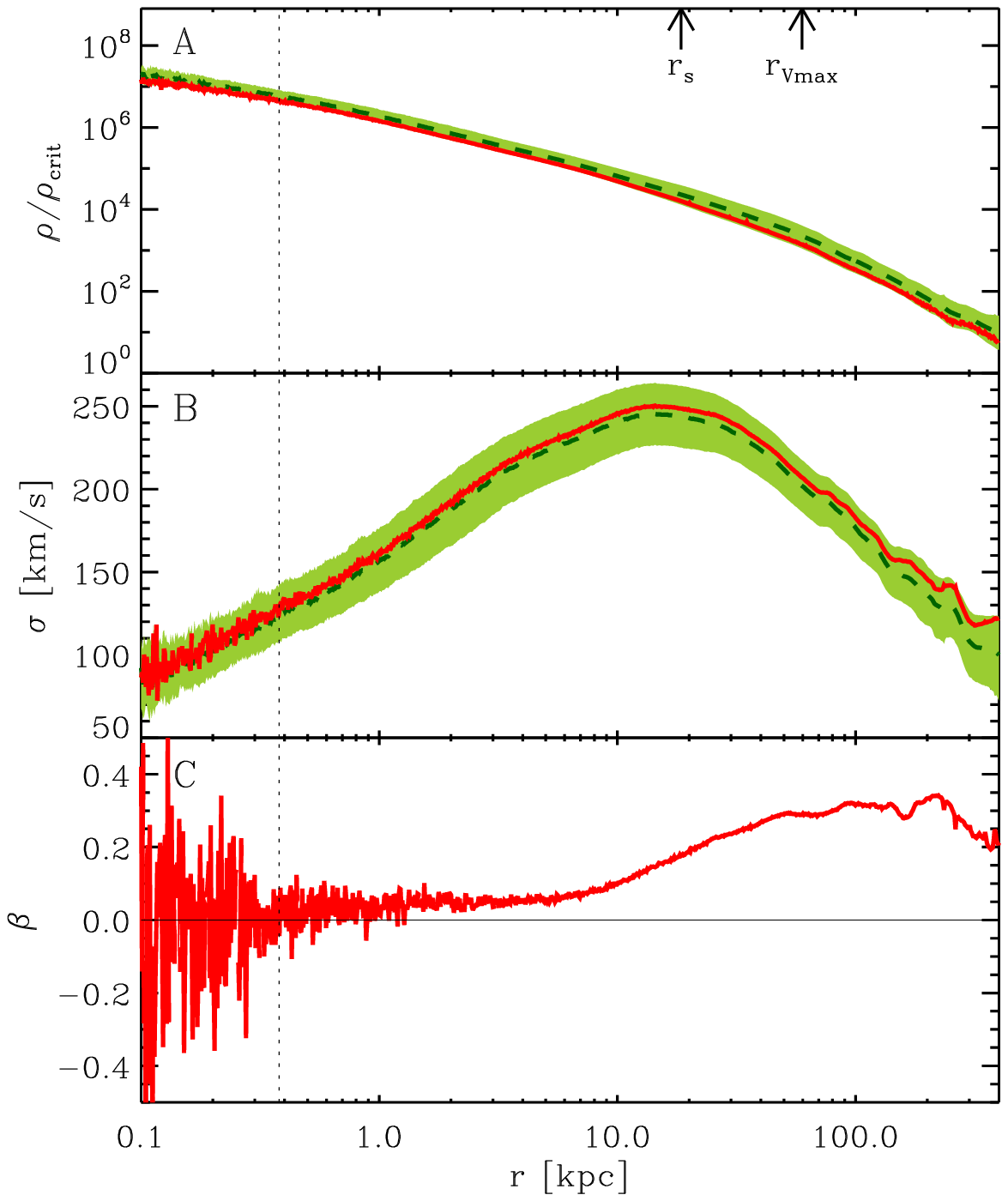}
\includegraphics[width=0.49\textwidth]{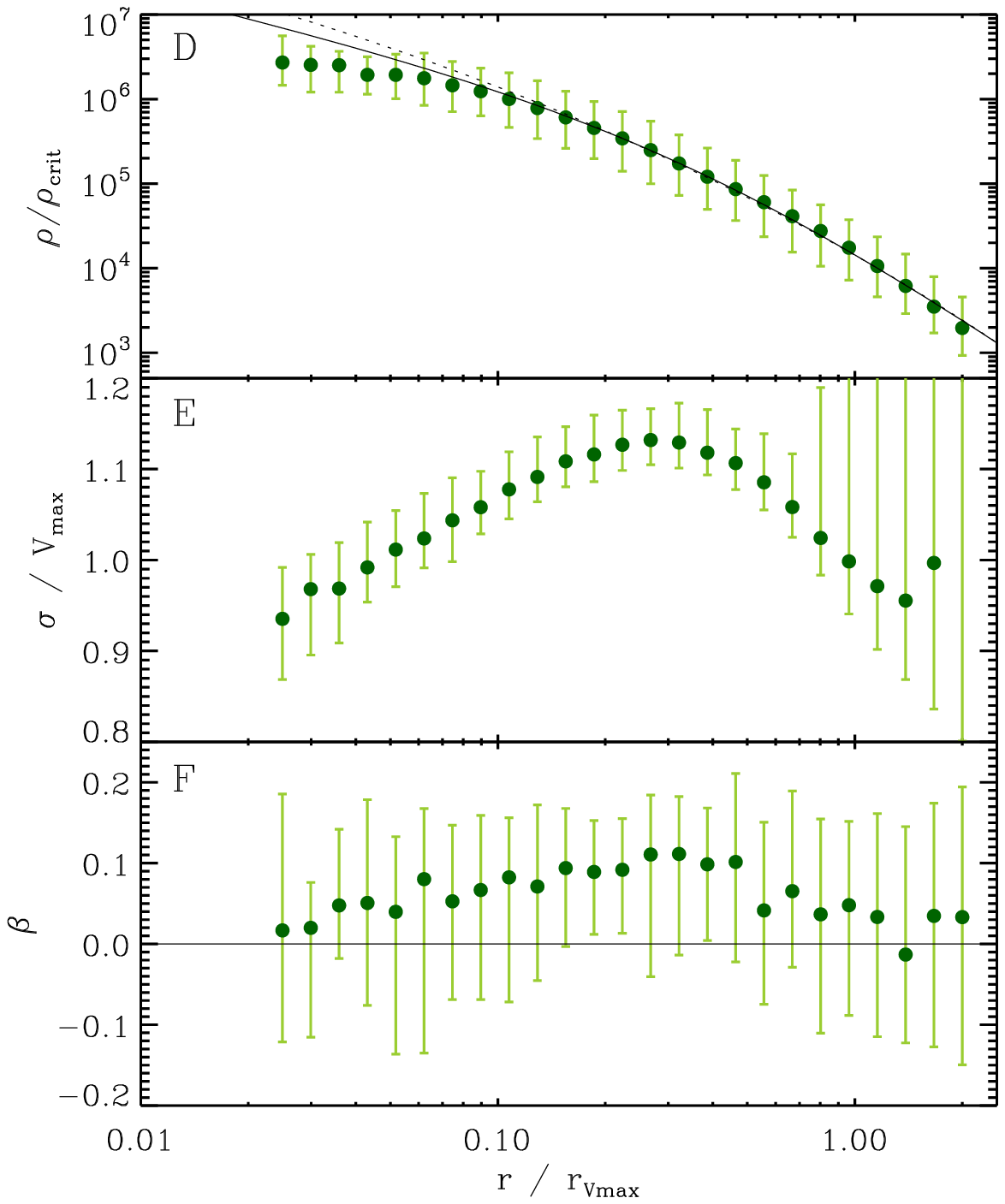}
\vspace*{-0.25in}
\caption{\captionfontsize {\it Left panel:} The radial dependence of
  the density (A), velocity dispersion (B), and the velocity
  anisotropy $\beta = 1 - \f{1}{2} \f{\sigma_\theta^2 +
    \sigma_\phi^2}{\sigma_r^2}$ (C) for the smooth host halo
  component.  The solid red line shows the values calculated from all
  particles in spherical shells. For the $\rho$ and $\sigma$ profiles
  we also show the median (dark green dashed line) and the 68\% region
  (light green shaded region) of the particle-by-particle quantities
  determined from the nearest 32 neighbors. The vertical dashed line
  indicates our estimate for the convergence radius of the density
  profile (380 pc). The shape of the $\sigma$ profile implies that
  Sommerfeld enhancement will preferentially brighten the very central
  region of the host halo at the expense of the surrounding region.
  The anisotropy profile clearly shows that the host halo is not
  isotropic, with a slight radial anisotropy persisting down to the
  convergence radius. {\it Right panel} (D-F): The same quantities
  averaged over the 100 most massive subhalos. The bullets indicate
  the median and the error bars the 68\% scatter around the median.
  In D and E we also plot the best-fitting NFW (solid) and Einasto
  ($\alpha$ fixed at 0.17, dotted) profiles. Note that the subhalo
  profiles should not be considered converged below $\sim 0.1\;
  \rVmax$.}
\label{fig:profiles}
\end{figure}

In Figure~\ref{fig:profiles} we present radial profiles of the density
$\rho$, velocity dispersion $\sigma$, and velocity anisotropy
parameter $\beta$ for the smooth host halo and averaged over the 100
most massive subhalos. We determine these profiles by first binning
all particles into equally spaced logarithmic radial shells, and then
calculating
\begin{eqnarray}
\rho(r) & = & \f{\sum_i m_i}{4\pi r^2 dr}, \\
\sigma_j^2(r) & = & \langle (v_j - \langle v_j \rangle_i )^2 \rangle_i, \\
\sigma^2(r) & = & \sigma_x^2(r) + \sigma_y^2(r) + \sigma_z^2(r), \\
\beta(r) & = & 1 - \f{1}{2} \f{\sigma_\theta^2(r) + \sigma_\phi^2(r)}{\sigma_r^2(r)},
\end{eqnarray} 
where the sum and averages (denoted by $\left< \right>_i$) are over
all particles in a given spherical shell. These profiles are indicated
by the solid red line for the host halo in the left panels of
Figure~\ref{fig:profiles}. For the $\rho(r)$ and $\sigma(r)$ profiles
we also show the median and 68\% interval of the distribution of
particle density $\rho_i$ and velocity dispersion $\sigma_i$, both
calculated from the 32 nearest neighboring particles. The two
estimates agree quite well with each other, with the slightly higher
(lower) median $\rho_i$ ($\sigma_i$) indicating a negative (positive)
skew of the distribution, presumably due to spherical averaging of a
triaxial mass distribution. Since 32 particles are not sufficient for
a good estimate of $\beta$, we don't show the distributions for the
velocity anisotropy.

The density profile of the \VLII\ host has been discussed in
\cite{S_Diemand2008} and we present it here merely for completeness.
Down to our convergence radius of $380$ pc it is well fit by a
generalized NFW profile,
\begin{equation}
\rho(r) = \f{\rho_s \; 2^{3-\gamma}}{(r/r_s)^\gamma (1+r/r_s)^{3-\gamma}},
\end{equation}
with a central slope of $\gamma=1.2$, but an
Einasto profile,
\begin{equation}
\rho(r) = \rho_s \exp{\left[-\f{\alpha}{2}\left( (r/r_s)^\alpha - 1 \right)\right]},
\end{equation}
with $\alpha = 0.167$ fits almost as well. The
velocity dispersion profile exhibits the central ``temperature
inversion'' typical of cold dark matter halos: $\sigma(r)$ peaks at
about the scale radius and decreases with decreasing radius
\cite{S_Diemand2004,S_Hoeft2004,S_Dehnen2005}. The $\beta(r)$ profile
shows the well established trend of a considerable amount of radial
anisotropy in the outskirts of the halo and decreasing towards the
center \cite{S_Diemand2004,S_Hoeft2004,S_Dehnen2005}. We find that
even at the convergence radius a slight amount of radial anisotropy
remains.

In the right panels of Figure~\ref{fig:profiles} we show the $\rho$,
$\sigma$, and $\beta$ profiles averaged over the 100 most massive
subhalos in our simulations. We scaled the radius of each subhalo by
its $\rVmax$ and calculated radial profiles using 30 equally spaced
logarithmic bins from $r/\rVmax = 0.01$ to 2. In each radial bin we
then determined the median and 68\% region of the distribution of
values over all 100 subhalos, rejecting bins containing less than 100
particles. We only plotted bins containing values from more than 10
subhalos. It is difficult to estimate a convergence radius for these
subhalos; the host halo convergence radius of 380 pc corresponds to
(0.05 - 0.5) $\rVmax$ for these 100 subhalos, so these average
profiles should not be considered converged below $\sim 0.1 \rVmax$.

The median density profile nicely follows the anticipated NFW-like
profile. We have overplotted the best-fitting NFW (solid line) and
Einasto (with fixed $\alpha=0.17$, dotted line) profiles. Clearly the
resolution is not good enough to allow a quantitative analysis of the
asymptotic central slope of the density profile, but it remains cuspy
as far down as we can resolve. The velocity dispersion profile looks
qualitatively the same as the host halo's, with a peak around $0.25
\rVmax$ and decreasing towards smaller radii. Not surprisingly,
subhalos are not isothermal. The strong increase in the $84^{\rm th}$
percentile of $\sigma$ at large radii is due to contamination by host
halo particles which artificially inflate $\sigma$. Although the
$\beta$ profile is quite noisy, it is clear that subhalos too exhibit
a slight radial anisotropy even down at the smallest radii that we can
access.


An important caveat to these findings is that our simulations
completely neglect the effects of baryons on the DM distribution. Gas
cooling, star formation, supernova feedback, and stellar dynamical
processes might alter both the DM density and velocity dispersion
profiles. In fact, not even the sign of these effects is clear at the
moment: adiabatic contraction generally leads to a steepening of the
central density profile \cite{S_Blumenthal1986,S_Gnedin2004}, but
dynamical friction acting on baryonic condensations tends to remove
the central cusp \cite{S_El-Zant2004,S_Romano-Diaz2008}. The velocity
dispersion of DM particles, on the other hand, is more likely to
increase in regions affected by baryonic processes. These complications
will be most important for the Galactic Center. The high mass-to-light
ratios observed in Galactic dwarf satellites \cite{S_Simon2007} indicate
that they are completely DM dominated and hence likely much less
affected by baryonic physics. Interactions with the Milky Way's
stellar and gaseous disk probably strip a significant fraction of mass
from some DM subhalos, but the dense central regions responsible for
most of the annihilation luminosity are relatively well protected.

\section{Detectability Calculation} \label{sec:detectability}

We calculated the annihilation flux directly from our simulations
following the procedure detailed in \cite{S_Kuhlen2008}, with one
important modification to account for the Sommerfeld enhancement. The
intensity in a given direction $(\theta,\phi)$ is now given by
\begin{equation}
  \mathcal{I}(\theta,\phi) = \mathcal{G} \int_{\rm los} \rho^2 S (\sigma;m_\chi,m_\phi,\alpha) d\ell,
\end{equation}
where $\mathcal{G}$ contains most of the particle physics dependence
(the particle mass, the high velocity annihilation cross section, and
the $\gamma$-ray spectrum per annihilation event) and
$S(\sigma;m_\chi,m_\phi,\alpha)$ is the Maxwell-Boltzmann-weighted
Sommerfeld enhancement factor at a velocity dispersion $\sigma$ for a
given particle physics model.  For discrete particles, with masses
$m_i$, distances $d_i$, densities $\rho_i$, and velocity dispersions
$\sigma_i$, this integral becomes a discrete sum over all particles in
a given map pixel,
\begin{equation}
\sum_i \f{\rho_i S(\sigma_i) m_i}{4\pi d_i^2}.
\end{equation}
We determined $\rho_i$ and $\sigma_i$ from the nearest 32 neighbors of
the $i$th particle, but have checked that our results do not change
significantly for 64 neighbors. We have implemented several additional
improvements over \cite{S_Kuhlen2008}: (a) we corrected a calculation
error in the conversion from simulation fluxes to gamma-ray counts
(the subhalo fluxes were a factor of two too large in
\cite{S_Kuhlen2008}), (b) we use a 15\% lower effective area ($\sim$
7,300 cm$^2$), as suggested by the performance of the LAT instrument
measured in orbit \cite{S_Johnson}, and (c) we switched to the
HEALPix\footnote{\texttt{http://healpix.jpl.nasa.gov/}} equal area
pixelization scheme, setting $N_{\rm side}=512$, which corresponds to
a solid angle per pixel of $\Delta \Omega = 4 \times 10^{-6}$ sterad,
comparable to the angular resolution of \Fermi's LAT detector above 3
GeV. We assumed a LAT exposure time equal to 0.153 of the time in
orbit, a combination of the $\sim 4\pi/5$ sr field of view, 90\% trigger
live time, and 15\% data acquisition outage during South Atlantic
Anomaly passages \cite{S_Johnson}. Our analysis was restricted to one
fiducial observer located at 8 kpc from the host halo center along the
intermediate axis of its density ellipsoid, and we refer the reader to
\cite{S_Kuhlen2008} for a discussion of the signal variance arising
from different observer locations.

Due to the finite resolution of our simulation the centers of all our
halos are artificially heated and less dense than they would be at
higher resolution. This results in central brightnesses that are lower
than would be expected for an NFW or Einasto density profile. We have
corrected the central flux from the host halo and all subhalos using
an analytical extrapolation of the density and velocity dispersion
profiles. We considered both an NFW ($\gamma=1$) profile
and an Einasto profile
with $\alpha=0.17$, which has been shown to fit Galactic-scale dark
matter halos \cite{S_Navarro2004}. These analytical profiles are
matched to the measured values of $\Vmax$ and $\rVmax$, which are
robustly determined for subhalos with more than 200 particles
($M>8\times 10^5 \msun$). The relations between $(\Vmax,\rVmax)$ and
$(\rho_s,r_s)$ are
\begin{equation}
\rVmax = f_r \; r_s \qquad \Vmax^2 = f_V \; 4\pi G \rho_s r_s^2,
\end{equation}
with $f_r \approx 2.163 \; (2.212)$ and $f_V \approx 0.865 \; (0.897)$
for the NFW (Einasto) profile. We use the spherical Jeans equation to
solve for the corresponding velocity dispersion profile, assuming
$\beta=0$. For a halo at a given distance we can then solve for a
Sommerfeld-enhanced surface brightness profile as a function of angle
from the halo center, average it over the angular resolution of our
maps ($\Delta\Omega = 4 \times 10^{-6}$ sterad), and use this to
correct our simulated maps. For the host halo we only correct pixels
within $\sim 2.7^\circ$ from the center, corresponding to the density
profile convergence radius of 380 pc. For the subhalos we ensure that
all pixels within the projected scale radius $r_s$ have a surface
brightness at least as high as the expectation from the analytical
extrapolation. These correction factors are typically not very large:
over all subhalos and all Sommerfeld models, the median (root mean
square) correction factor for the central pixel is 2.2 (5.0). 

Note that we neglect the possible enhancement of a subhalo's
luminosity arising from additional clumpy substructure below our
simulation's resolution limit. The magnitude of this so-called
``substructure boost factor'' depends on uncertain extrapolations of
the abundance, distribution, and internal properties of the low mass
subhalos, and will not be significantly increased by the Sommerfeld
effect due to its saturation at low velocities. While this boost
typically doesn't affect the central surface brightness very much, it
may somewhat increase the angular extent of a given subhalo's signal.

The annihilation signal from individual subhalos must compete with a
number of diffuse $\gamma$-ray backgrounds, of both astrophysical and
DM annihilation origin. At low Galactic latitudes the dominant
astrophysical background arises arises from the interaction of high
energy cosmic rays with interstellar gas (pion decay and
bremsstrahlung) and radiation fields (inverse Compton). This
background has been measured by the EGRET instrument aboard the
Compton satellite at 0.5 degree resolution out to $\sim 30$ GeV
\cite{S_Hunter1997}. An improved measurement of the spectral and
angular properties of this background is one of the goals of the
\Fermi\ mission, and preliminary data at intermediate Galactic
latitudes have already been presented by the Fermi collaboration
\cite{S_Moskalenko2009}. Here we employ a theoretical model of this
background (the GALPROP ``conventional'' model
\cite{S_Strong1998,S_Moskalenko2002}), which matches the EGRET and
preliminary Fermi measurements with sufficient accuracy for our
purposes. At high Galactic latitudes ($|b| \gta 20^\circ$), an
isotropic extragalactic $\gamma$-ray background from unresolved
blazars may become dominant. We include such a background with an
intensity and spectrum as measured by EGRET \cite{S_Sreekumar1998}.
Note that this is probably an overestimate, since \Fermi\ will likely
resolve some fraction of this background into point sources. We do not
include the contributions from astrophysical foreground and background
point sources, but we have checked that none of the
Sommerfeld-enhanced subhalos would have been detected by EGRET,
assuming a point-source sensitivity of $2 \times 10^{-7} \gamma$
cm$^2$ s$^{-1}$ from 0.1 to 30 GeV \cite{S_Hartman1999}.

In addition to these astrophysical backgrounds, individually
detectable subhalos must compete with the diffuse background from DM
annihilation in the smooth host halo and from the population of
individually undetectable subhalos (see Section~\ref{sec:unresolved})
\cite{S_Pieri2008,S_Kuhlen2008}. In models without Sommerfeld
enhancement and for $S \sim 1/v$ models, these DM annihilation
backgrounds are negligible compared to the astrophysical backgrounds
(although they themselves constitute a signal worth searching for). In
resonant $S \sim 1/v^2$ models with low saturation velocity (e.g.
LS-3, LS-4, MPV-1b, MPV-2b), however, the unresolved subhalo flux
becomes the dominant background and must be accounted for. For
completeness we have included all four components in the detectability
calculation in all cases.

\begin{figure}
\centering
\includegraphics[width=0.5\columnwidth]{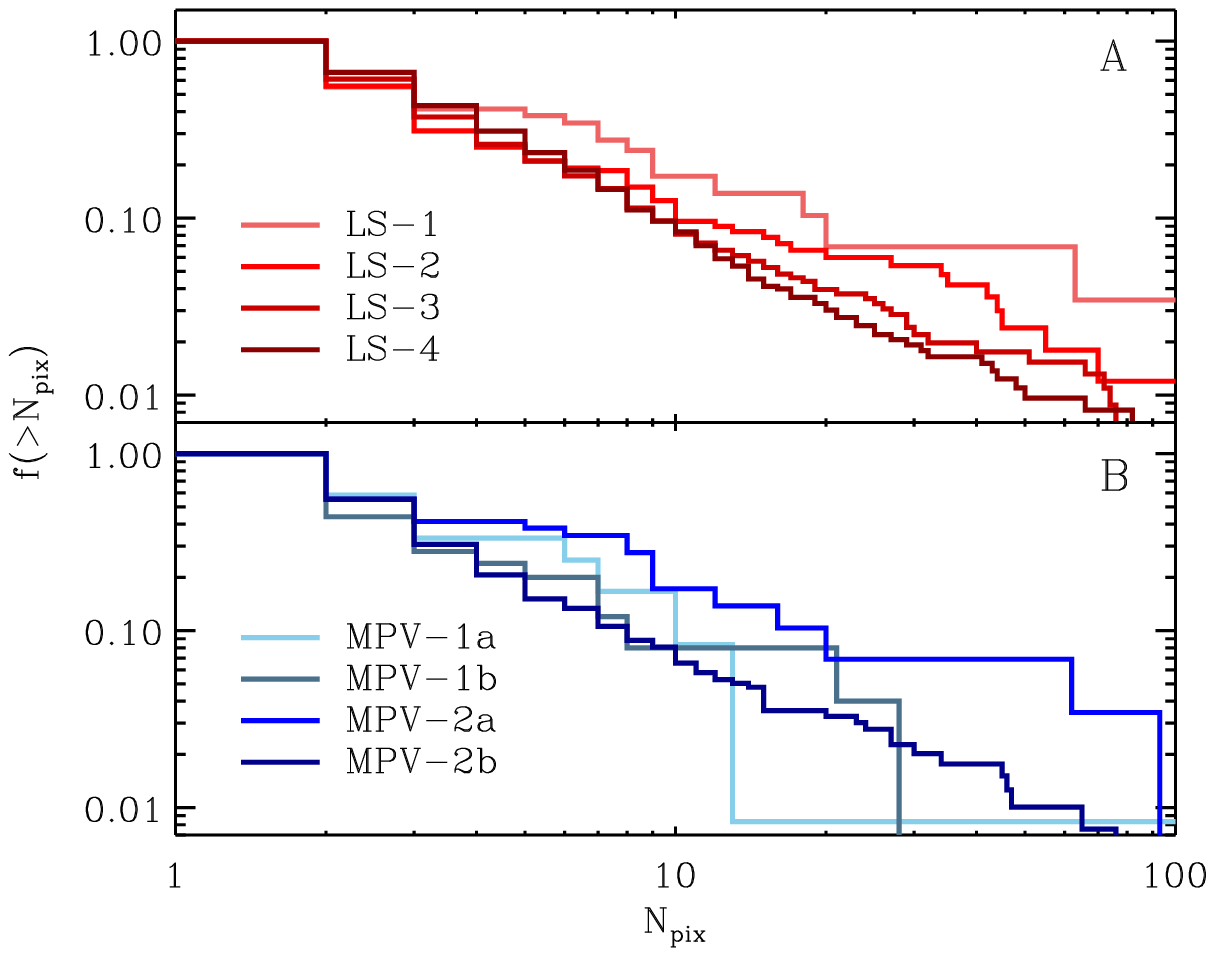}
\caption{\captionfontsize Cumulative distribution of the angular size
  of the detectable subhalos. We plot the fraction of S/N$ >
  5$ subhalos with more than $N_{\rm pix}$ pixels exceeding the
  \Fermi\ detection threshold after 5 years in orbit, for the LS
  models in panel A, and for the MPV models in panel B. }
\label{fig:Npix}
\end{figure}

Subhalo detectability for \Fermi\ is assessed as follows. First we
calculate a signal-to-noise ratio ($S/N$) \textit{per pixel} by
dividing the number of source photons arising in the subhalo map by
the square root of the number of photons in the background map. Next
we select all pixels with $S/N>1$ and identify contiguous regions,
which we associate with individual subhalos based on proximity of the
brightest pixel with a subhalo center. If more than one subhalo center
coincides with a given brightest pixel, we pick the subhalo with the
larger expected surface brightness ($L/r_s^2 \sim \Vmax^4/\rVmax^3$).
For each of these contiguous regions we then calculate a subhalo
detection significance
\begin{equation}
S/N = \f{N_s}{\sqrt{N_b}},
\end{equation}
where $N_s$ and $N_b$ are the total number of source and background
$\gamma$-rays over the contiguous region. This definition is a good
proxy for detection significance under the assumption that an estimate
of the background can be subtracted out and only Poisson fluctuations
remain.  Note that it is not the uncertainty of the flux itself (which
would be $N_s/\sqrt{N_s+N_b}$), but instead an estimate of the
significance of having detected a departure from a smooth background.

In Figure~\ref{fig:Npix} we show the cumulative distribution of the
angular size (the number of pixels exceeding the \Fermi\ detection
threshold after 5 years in orbit) of the detectable subhalos. Although
models with stronger Sommerfeld enhancement result in smaller
detectable regions, the majority of all subhalos would still be
resolved sources for \Fermi. This effect raises the possibility of
future observations being able to discriminate between different
Sommerfeld-enhanced models.

\section{Diffuse Flux from Unresolved Subhalos}\label{sec:unresolved}

\begin{figure}
\centering
\includegraphics[width=0.5\columnwidth]{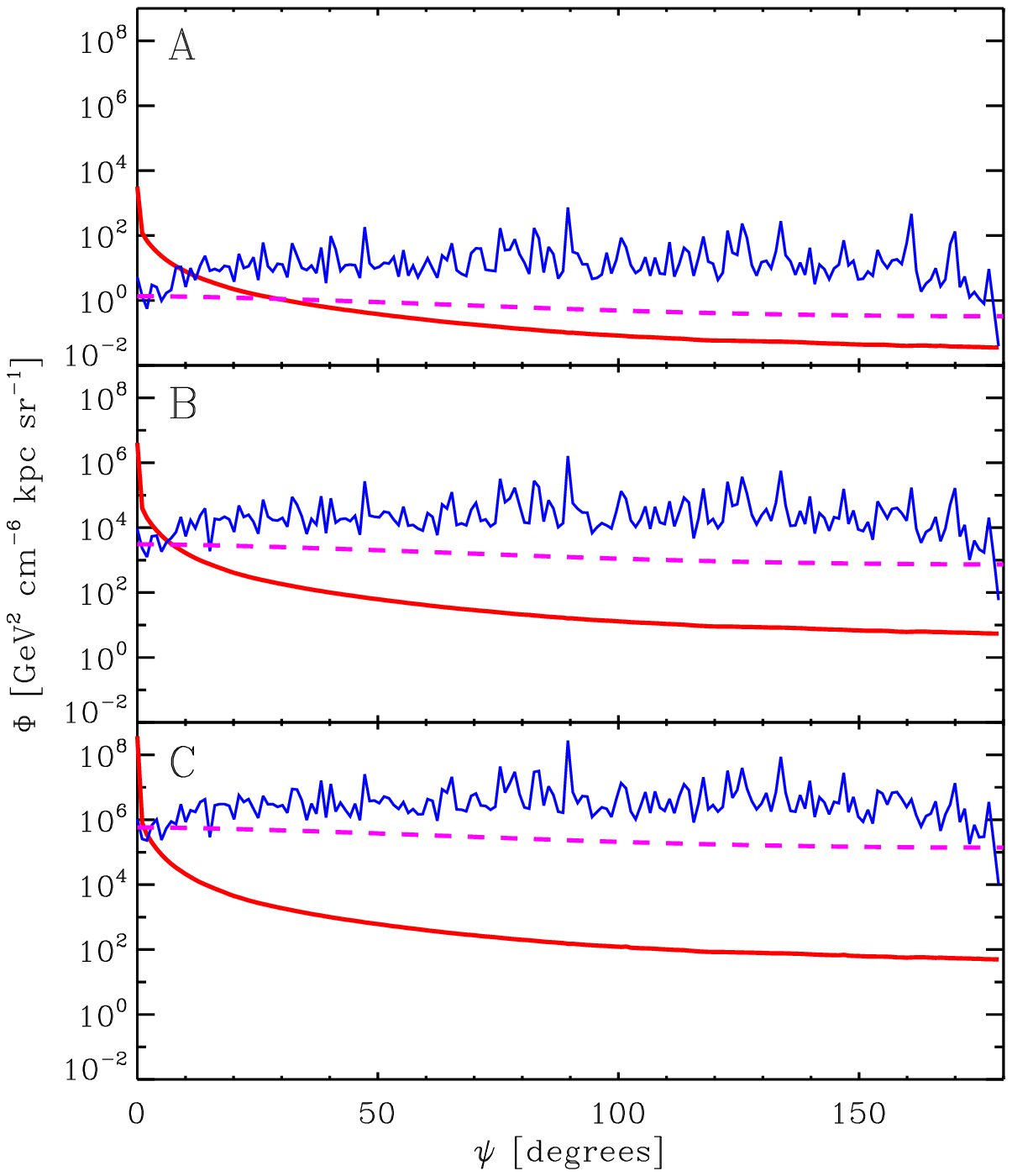}
\caption{\captionfontsize The annihilation intensity as a function of
  angle $\psi$ from the Galactic Center, for the host halo (red),
  individual subhalos (blue), and unresolved subhalos (magenta
  dashed). A: No Sommerfeld enhancement. B: Model MPV-2a ($S \sim
  1/v$, high $v_{\rm sat}$). C: Model LS-4 ($S \sim 1/v^2$, low
  $v_{\rm sat}$).  For the host halo and unresolved components we plot
  the \textit{mean}, for the individual subhalo profile the
  \textit{maximum} intensity over all pixels in a given $\psi$ bin.
  Individual subhalos outshine the diffuse unresolved subhalo
  background, even in the strongly Sommerfeld-enhanced case.}
\label{fig:flux_profile}
\end{figure}

For a typical CDM power spectrum of density fluctuations one would
expect dark matter clumps on scales all the way down to a cutoff set
by collisional damping and free streaming in the early universe
\cite{S_Green2005,S_Loeb2005}. For WIMP dark matter, typical cut-off
masses are $\mcut = 10^{-12}$ to $10^{-4} \msun$
\cite{S_Profumo2006,S_Bringmann2009}, some 10 to 20 orders of
magnitude below \VLII's mass resolution. In this case the Galactic
dark matter halo might host an enormous number ($10^{10} - 10^{22}$,
see \cite{S_Kuhlen2008}) of small mass subhalos, whose combined
annihilation signal could result in a sizeable $\gamma$-ray
background. If the Sommerfeld enhancement didn't saturate at a finite
velocity, this background would easily outshine any other Galactic
$\gamma$-ray signal. Even with saturation one must ask whether the
Sommerfeld-enhanced annihilation background from this population of
unresolved subhalos would outshine individual subhalos.

In order to address this question, we have extended the analytical
model of \cite{S_Kuhlen2008} to allow for Sommerfeld enhancement. The
overall intensity of this background depends sensitively on a number
of uncertain parameters governing the subhalo population as a whole
(slope and normalization of the mass function, their dependence on
distance to the host center) and the mass dependent properties of
individual subhalos (density and velocity dispersion profiles,
concentrations). Although the model is calibrated to numerical
simulations at the high mass end, it relies on an extrapolation over
many orders of magnitude in mass below the simulation's resolution
limit that is very uncertain.

Our model employs a radially anti-biased subhalo mass function
\begin{equation}
\f{dn(M,r)}{dM} = 1.05 \times 10^{-8} \msun^{-1} \; {\rm kpc}^{-3} \left( \f{M}{10^6 \msun} \right)^{-2} \left( 1 + \f{r}{18.5\;{\rm kpc}}\right)^{-2}
\end{equation}
with a low mass cutoff of $\mcut = 10^{-6} \msun$, and assumes an Einasto
density and velocity dispersion profile with a concentration-mass
relation according to \cite{S_Bullock2001} and a radial dependence of
\begin{equation}
c(M,r) = c^{\rm B01}(M) \left( \f{r}{400\;{\rm kpc}} \right)^{-0.286}.
\end{equation}
We refer the reader to the Appendix of \cite{S_Kuhlen2008} for details
about the calculation. In Figure~\ref{fig:flux_profile} we show the
resulting azimuthally averaged intensity as a function of angle from
the Galactic Center, and compare it to the smooth host halo signal and
the peaks due to individual subhalos. While the unresolved subhalo
background is brighter than the smooth halo component everywhere but
in the very center, individual subhalos have higher central surface
brightnesses and can easily outshine it. This holds true in all
Sommerfeld models that we have considered.

\section{Local Luminosity Boost From Substructure}

\begin{figure*}
\centering
\includegraphics[width=0.5\columnwidth]{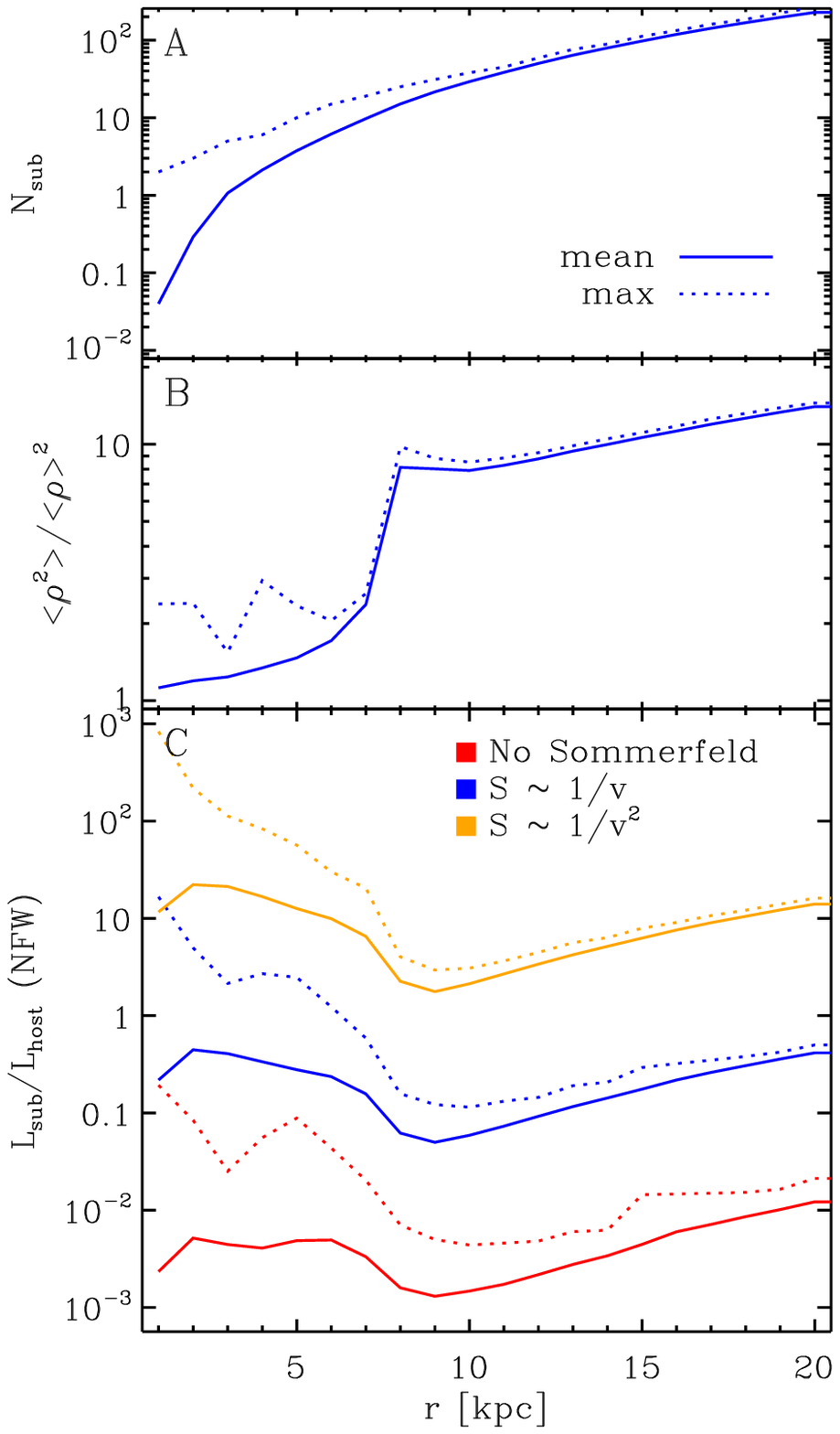}
\caption{\captionfontsize The effect of substructure on the local
  annihilation rate. A: The mean and maximum number of \VLII\ subhalos
  inside 100 randomly placed spheres 8 kpc from the halo center versus
  the radius of these sample spheres. The mean subhalo occupancy
  becomes unity at $r=3$ kpc. B: The density ``clumping factor''
  $\langle \rho^2 \rangle / \langle \rho \rangle^2$ over the 100
  sample spheres.  C: The ratio of the subhalo to host halo
  contributions to the annihilation luminosity for three
  representative Sommerfeld models (none, $1/v$, and $1/v^2$). Only
  subhalos resolved in our simulation are accounted for.}
\label{fig:local_profiles}
\end{figure*}

Here we assess the role that nearby subhalos play for the
Sommerfeld-enhanced production of high energy electrons and positrons.
Because these energetic particles lose energy as they diffuse through
the Galactic magnetic field, only those that are produced within a few
kpc of Earth are of interest.

We considered 100 spheres of radius 20 kpc, with randomly positioned
centers 8 kpc from the host halo center. For each of these spheres we
determined the cumulative number of subhalos $N_{\rm sub}$, a
``clumping factor'', defined as $\langle \rho^2 \rangle / \langle \rho
\rangle^2$, and the ratio of total subhalo to host halo luminosity as
a function of enclosed radius in the sphere. For the subhalo
luminosity we used the analytical NFW estimate, as explained in
Section~\ref{sec:detectability}. Figure~\ref{fig:local_profiles} shows
the mean and maximum values of these quantities over all 100 sample
spheres. The low local subhalo abundance is reflected in a small mean
$N_{\rm sub}$ within a few kpc of the Sun. Only 3 of the 100 sample
spheres have any subhalos within 1 kpc of their center. The mean
subhalo occupancy becomes unity at 3 kpc, but one has to go out to 7
kpc before every single sphere contains at least one subhalo. The
clumping factor captures the enhancement of the annihilation
luminosity compared to a homogeneous density background. It has
contributions from the overall density stratification, from the
Poisson noise of the density estimator, and from unbound and bound
substructure. The sharp rise towards 8kpc is due to the cuspy nature
of the host halo density profile. The bottom panel of
Figure~\ref{fig:local_profiles} shows that without Sommerfeld
enhancements the subhalos do not contribute significantly to the local
annihilation luminosity. In a typical $S \sim 1/v$ model, however,
subhalos contribute on average about half as much as the host halo,
and in rare cases 5 times more. For models on resonance ($S \sim
1/v^2$), the subhalos completely dominate the host halo, and provide
on average 20 times as much luminosity as the host halo.  Again we
have neglected the possible additional contribution from subhalos
below our simulation's resolution limit.

{}

\end{document}